\documentclass[twocolumn,showpacs,preprintnumbers,amsmath,amssymb]{revtex4}

\usepackage{graphicx}
\usepackage{dcolumn}
\usepackage{bm}

\begin{document}


\title{Enhanced annealing effect in an oxygen atmosphere on $\mathbf{Ga_{1-x}Mn_{x}As}$}

\author{M. Malfait}
\author{J. Vanacken}
\author{V. V. Moshchalkov}%

\affiliation{%
Pulsed Field Group, Laboratory of Solid State Physics and
Magnetism, K.U.Leuven, Celestijnenlaan 200D, 3001 Leuven, Belgium
}%

\author{W. Van Roy}
\author{G. Borghs}
\affiliation{ IMEC, Kapeldreef 75, 3001 Leuven, Belgium}%


\begin{abstract}
We report on in-situ resistivity measurements on $\rm
Ga_{1-x}Mn_{x}As$ during post-growth annealing in different
atmospheres. A drop in the resistivity is observed when the $\rm
Ga_{1-x}Mn_{x}As$ is exposed to oxygen, which indicates that the
passivation of Mn interstitials ($\rm Mn_I$) at the free surface
occurs through oxidation. The presence of oxygen can therefore be
an important annealing condition for the optimization of $\rm
Ga_{1-x}Mn_{x}As$ thin films, all the more since the oxidation
appears to be limited to the sample surface. Annealing in an
oxygen-free atmosphere leads to an increase in the resistivity
indicating a second annealing mechanism besides the out-diffusion
of $\rm Mn_I$. According to our magnetization and Hall effect
data, this mechanism reduces the amount of magnetically and
electrically active Mn atoms.
\end{abstract}

\maketitle


The combination of ferromagnetism with the versatile
semiconducting properties in III-V ferromagnetic semiconductors
such as $\rm Ga_{1-x}Mn_{x}As$ makes them promising for future
spintronics applications, in which the spin degree of freedom is
used to process and transfer information. The amount of Mn, that
has to be incorporated to obtain ferromagnetic $\rm
Ga_{1-x}Mn_{x}As$ epilayers, is far above the equilibrium
solubility limit. Therefore these films have to be grown at low
temperatures using molecular beam epitaxy  (MBE) \cite{Ohno}. When
Mn ions are incorporated into GaAs, they do not only introduce
magnetic moments but also act as acceptors, providing holes that
mediate the interaction between the localized Mn spins. Obtaining
higher values for the Curie temperature $\rm T_C$ is therefore not
only a matter of increasing the Mn content, but also of achieving
a higher free carrier density. The low growth temperatures,
however, lead to a high density of Mn interstitials ($\rm Mn_I$)
and As antisite defects ($\rm As_{Ga}$) which both act as double
donors \cite{Sanvito,Maca,Erwin}, and therefore compensate a
fraction of the holes generated by the substitutional $\rm
Mn_{Ga}$ acceptor. Several authors have reported an enhancement of
the hole concentration and $\rm T_C$ upon after growth annealing
at temperatures close to the growth temperature
\cite{Hayashi,Potashnik1,Potashnik2,Kuryliszyn,Edmonds1,Ku,Chiba,Wang}.
The best results so far were obtained for annealing temperatures
$\rm T_A$ just below the growth temperature \cite{Edmonds1,Wang},
while long annealing times and higher temperatures result in a
reduction of $\rm T_C$
\cite{Hayashi,Potashnik1,Potashnik2,Kuryliszyn,VanEsch}. The
optimization of $\rm T_C$ can be done in a controlled way by
monitoring the resistance while annealing \cite{Edmonds1}, as the
high temperature resistivity is correlated with the hole
concentration, and therefore with $\rm T_C$. From ion channeling
experiments Yu et al. showed strong evidence that this increase of
$\rm T_C$ is related to the removal of $\rm Mn_I$ atoms \cite{Yu},
while Edmonds et al. recently identified the underlying mechanism
with the outdiffusion of the highly mobile $\rm Mn_I$ to the free
surface \cite{Edmonds2}. It was already suggested by Edmonds et
al. that at the surface the interstitial Mn atoms may be
passivated by oxidation \cite{Edmonds2}. This mechanism is
supported by the capping-induced suppression of the annealing
effects \cite{Chiba,Stone}. In this paper we present evidence that
the presence of oxygen is an important parameter for the
optimization of $\rm T_C$ through annealing, indicating that the
passivation of $\rm Mn_I$ indeed occurs through oxidation.

$\rm Ga_{1-x}Mn_{x}As$ films with various Mn content (0.03 $\leq$
x $\leq$ 0.08) were deposited by standard low-temperature
molecular beam epitaxy on semi-insulating epi-ready (001) GaAs
substrates. The growth was performed with a nearly stoichiometric
$\rm As_2$ flux at temperatures typically $\approx 15^{\circ}$C
below the Mn segregation limit, on a 100 nm high-temperature GaAs
buffer layer grown under standard conditions, followed by a low
temperature GaAs buffer of a similar thickness. The growth was
monitored in situ by reflection high energy electron diffraction
(RHEED), which showed a clear ($1\times2$) reconstruction. The
structural quality and the sample thickness of about 40 nm were
checked with XRD measurements. A part of the wafer was then
chemically etched into Hall bars using photo-lithography. A small
variation of about 3\% in the room temperature resistivity was
observed in samples that were taken from the same wafer.
Post-growth annealing was performed in a tube furnace, which
allows a well controlled atmosphere, consisting of vacuum or a
specific gas, for the annealing procedure. The resistance was
measured in-situ in a four probe configuration. Magnetization
measurements were performed with vibrating sample magnetometer
(VSM) on unprocessed samples with a typical size $\rm \sim
30~mm^2$.

\begin{figure}
\includegraphics{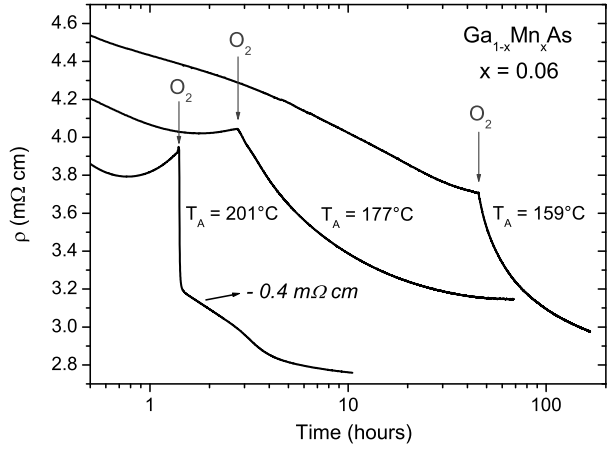}
\caption{\label{Fig1} Resistivity versus annealing time for 40 nm
thick $\rm Ga_{1-x}Mn_{x}As$ (x = 0.06) films at 159$^{\circ}$C,
177$^{\circ}$C and 201$^{\circ}$C, measured in a four probe
configuration. The annealing is initially performed in a forming
gas atmosphere ($\rm N_2 + 1\%~H_2$). The arrows indicate when the
samples were exposed to oxygen. The curve at 201$^{\circ}$C is
shifted by -0.4 m$\rm \Omega$ cm for clarity.}
\end{figure}

To investigate the role of oxygen in the annealing process, we
have subjected the samples to an after-growth annealing procedure
in which the atmosphere was initially oxygen-free and after a well
defined annealing time the epilayers were exposed to oxygen. To
ensure the oxygen-free initial atmosphere, the quartz tube
containing the sample was pumped to a vacuum of the order of
$10^{-6}$ mbar, flushed out and filled with forming gas consisting
of 99\% $\rm N_2$ gas and 1\% $\rm H_2$ gas. The tube was then
sealed, while filled with forming gas at a slight overpressure
($\approx$ 0.1 bar). Within about 20 minutes the annealing
temperature is stabilized with little overshoot ($\leq
1^{\circ}$C). After a well controlled time, the tube containing
the sample is exposed to a gentle $\rm O_2$ gas flow, typically
for 5 to 10 minutes, maintaining a constant temperature and
ensuring an oxygen pressure of the order of 1 bar. During this
entire procedure the resistance is continuously monitored.
Fig.~\ref{Fig1} shows the resistivity as a function of time for
$\rm Ga_{1-x}Mn_{x}As$ (x = 0.06) at 159$^{\circ}$C,
177$^{\circ}$C and 201$^{\circ}$C. The resistivity initially
decreases with a rate that diminishes with time (note that the
time in Fig.~\ref{Fig1} is plotted on a logarithmic scale), and
then rises again for $\rm T_A$ = 177$^{\circ}$C and
201$^{\circ}$C, which will be discussed later. Upon exposure to
oxygen a sudden resistivity drop occurs, which is very sharp for
$\rm T_A$ = 201$^{\circ}$C, and shows diffusion-like behavior
similar to that reported by Edmonds et al.
\cite{Edmonds1,Edmonds2}. A similar drop in resistivity occurs
when $\rm N_2$ gas is used as initial atmosphere or for samples
with a different Mn content. These results clearly show that the
presence of oxygen enhances the resistivity reduction upon
annealing, which corroborates the migration of hole-compensating
interstitials to the free surface during the annealing process
\cite{Edmonds2}, where they are passivated through oxidation. This
reduction of compensating defects results in the observed increase
in hole density and conductivity. The $\rm Ga_{1-x}Mn_{x}As$
epilayer itself does not appear to suffer from oxidation, since
this would lead to an increase in the resistivity. This is
confirmed by low angle XRD measurements on a 235 nm $\rm
Ga_{1-x}Mn_{x}As$ (x = 0.07) film, which indicate a natural oxide
layer with a thickness of 2 nm, while after annealing in $\rm O_2$
gas ($\approx 200^{\circ}$C, 80 hours) this thickness only
increased to about 5 nm.

\begin{figure}
\includegraphics{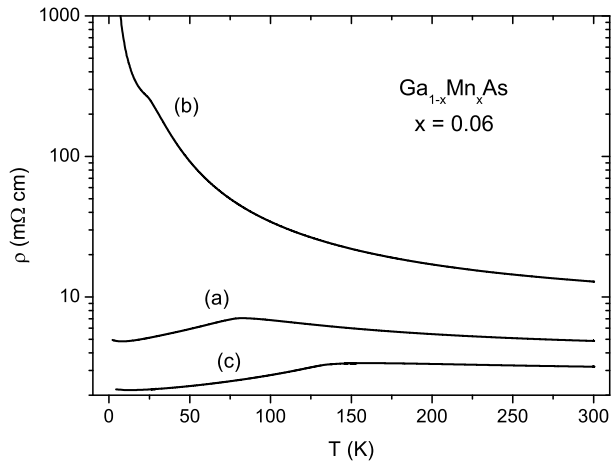}
\caption{\label{Fig2} Temperature dependence of the resistivity of
$\rm Ga_{1-x}Mn_{x}As$ (x = 0.06) films (a) as-grown, (b) after
annealing with only forming gas (198$^{\circ}$C, 16 hours),  (c)
after annealing in an oxygen atmosphere (190$^{\circ}$C, 70
hours).}
\end{figure}

Fig.~\ref{Fig2} shows the resistivity versus temperature for a
sample with x = 0.06, both as-grown (curve a) and after annealing
at $\approx 200^{\circ}$C for 60 hours in an $\rm O_2$ atmosphere
(curve c). Both curves show typical behavior for metallic $\rm
Ga_{1-x}Mn_{x}As$, with a peak at $\rm T_p$ = 82 K for the
as-grown sample, and $\rm T_p$ = 162 K for the oxygen-annealed
one. The peak temperature gives a good estimate of the Curie
temperature for the as grown sample. The direct measurement of
$\rm T_C$ with vibrating sample magnetometry gives $\rm T_C$ = 81
K. However, for the annealed sample $\rm T_p$ is found to
overestimate the measured $\rm T_C$ = 133 K by 29 K, which is a
large deviation, even considering the broadness of the resistivity
peak.

\begin{figure}
\includegraphics{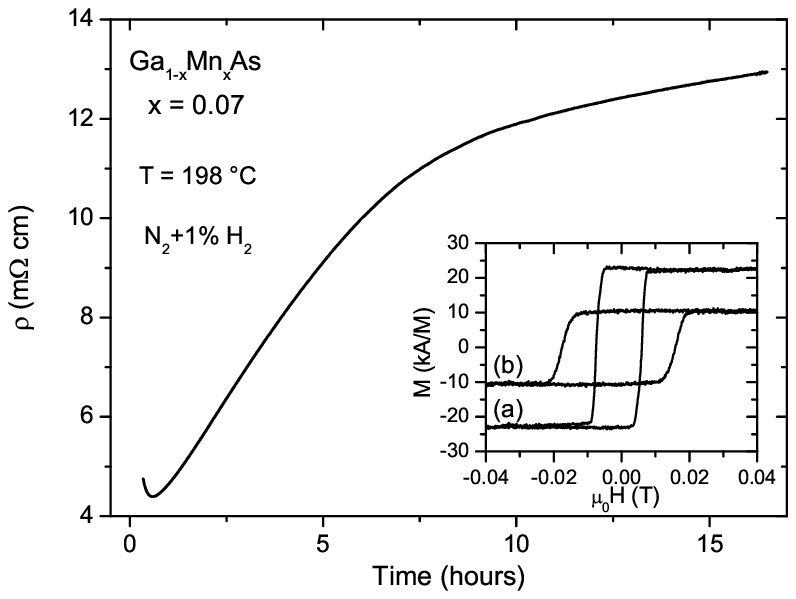}
\caption{\label{Fig3} Resistivity versus annealing time for a $\rm
Ga_{1-x}Mn_{x}As$ (x = 0.07) film in a forming gas atmosphere at
198$^{\circ}$C. The inset shows hysteresis loops measured at 5 K
for $\rm Ga_{1-x}Mn_{x}As$ (x = 0.07): (a) as-grown, measured
along the [100] easy axis; (b) after annealing in a forming gas
atmosphere (199$^{\circ}$C, 25 hours), measured along the [001]
easy axis.}
\end{figure}

To investigate further the increase in resistivity observed at
$\rm T_A$ =  177$^{\circ}$C and 201$^{\circ}$C (before exposure to
$\rm O_2$), the same annealing procedure was repeated without
oxygen, thus in a forming gas atmosphere. The resistivity of $\rm
Ga_{1-x}Mn_{x}As$ (x = 0.07) as a function of time at
198$^{\circ}$C is shown in Fig.~\ref{Fig3}. The resistivity
initially drops, but then quickly increases with a rate that
decreases with time. A similar effect is observed for samples with
various Mn content and the slowly increasing resistivity was
established for annealing times exceeding 100 hours. It is
unlikely that the increase in resistivity is caused by the
passivation of the $\rm Ga_{1-x}Mn_{x}As$ layer due to
hydrogenation as described recently by Brandt et al.
\cite{Brandt}, as a similar resistivity curve was observed when
annealing in a $\rm N_2$ gas atmosphere and in vacuum. The
resistivity increase is even observed at $\rm T_A$ =
158$^{\circ}$C when annealing $\rm Ga_{1-x}Mn_{x}As$ (x = 0.04 and
x = 0.08) in vacuum, albeit only by about 20\% after annealing for
120 hours. These results are in agreement with the increased
resistance observed for long annealing times and higher
temperatures
\cite{Hayashi,Potashnik1,Potashnik2,Kuryliszyn,VanEsch}. The
resistivity versus temperature for $\rm Ga_{1-x}Mn_{x}As$ (x =
0.06) after annealing in forming gas is plotted in fig. 2 as curve
(b) and shows semiconducting behavior with a slight deviation
around 25 K. To determine the carrier type and concentration, Hall
effect measurements in DC magnetic fields up to 12 T were
performed at 300K, which is far above $\rm T_C$ so no significant
paramagnetic contribution through the anomalous Hall effect was
detected. The Hall data revealed positive holes as carriers with a
concentration of $4.6 \times 10^{19}$ cm$^{-3}$ for $\rm
Ga_{1-x}Mn_{x}As$ (x = 0.06) annealed in forming gas, which is
almost an order of magnitude less than the hole concentration p =
$2.0 \times 10^{20}$ cm$^{-3}$ found for the as-grown sample from
Hall measurements at 5 K. The latter value for p was obtained by
taking the magnetoresistance contribution to the anomalous Hall
effect into account, by assuming the anomalous Hall coefficient
$R_A \propto \rho^2$.

Magnetization measurements performed on $\rm Ga_{1-x}Mn_{x}As$ (x
= 0.07) show that the easy axes for as-grown samples are [100] and
[010] in-plane, while after a similar annealing treatment in
forming gas the easy axis is shifted to the out-of-plane [001]
axis, as expected for $\rm Ga_{1-x}Mn_{x}As$ with such a low hole
concentration \cite{Dietl,Sawicki}. When measuring along the easy
axis, the magnetic moment shows no significant increase when the
magnetic field is raised from 20 mT to 0.5 T, indicating that the
samples are in a nearly single domain state at remanence.
Macroscopic single domains in $\rm Ga_{1-x}Mn_{x}As$ have been
established with scanning SQUID microscopy \cite{Fukumura},
scanning Hall probe microscopy \cite{Welp} and planar Hall effect
measurements \cite{Tang}. Therefore the remanent magnetization
$\rm M_{rem}$ along the easy axis is a good measure of the
saturation magnetization of the $\rm Ga_{1-x}Mn_{x}As$ layer. As
can be seen from the inset of Fig.~\ref{Fig3}, $\rm M_{rem}$ has
decreased by more than 50 \% as a consequence of the annealing
procedure.

These results show that there is a second mechanism besides the
diffusion of $\rm Mn_I$. This mechanism appears to diminish the
amount of ferromagnetically coupled Mn ions, as $\rm M_{rem}$ has
decreased, and since the carrier concentration is strongly
reduced, this may be due to the removal of Mn from electrically
active Ga sites. The fact that the remaining carriers are still
found to be p-type indicates that $\rm Mn_{Ga}$ ions do not simply
move to interstitial positions, as an excess of $\rm Mn_I$ would
lead to electrons as carriers. The activation energy for the
removal of $\rm Mn_{Ga}$ will be much higher than that for the
out-diffusion of $\rm Mn_I$, which therefore dominates in an $\rm
O_2$ atmosphere. When no oxygen is available, no passivation at
the surface can occur, but the relatively mobile $\rm Mn_I$ are
still free to migrate through the epilayer. An additional
mechanism could be the formation of Mn-As complexes, possibly with
$\rm Mn_I$ as an intermediate state. MnAs inclusions were
previously observed after annealing at high temperatures
(600$^{\circ}$C) \cite{DeBoeck}. The initial decrease in
resistivity upon annealing as seen in Fig.~\ref{Fig1} and
\ref{Fig3} may be due to the passivation of some $\rm Mn_I$ at the
natural oxide layer at the sample surface and the limited
diffusion of $\rm Mn_I$ to the substrate.

In summary,  exposure of $\rm Ga_{1-x}Mn_{x}As$ to oxygen during
annealing causes a substantial drop in the resistivity, which
indicates that the passivation of $\rm Mn_I$ at the free surface
occurs through oxidation. The presence of oxygen can therefore be
an important post-growth annealing condition for the optimization
of this system. Annealing in an oxygen-free atmosphere leads to an
increase in the resistivity indicating a second annealing
mechanism besides the out-diffusion of $\rm Mn_I$, that reduces
the amount of electrically and magnetically active Mn atoms.


\begin{acknowledgments}
This work was supported by the Fund for Scientific Research -
Flanders (Belgium) (F.W.O-V), the Research Fund K.U.Leuven,
GOA/2004/02 and IUAP programs. The IMEC-team acknowledges the
financial support of EC-project FENIKS G5RD-CT-2001-00535. W.V.R.
acknowledges support as a Postdoctoral Fellow of the F.W.O-V. The
authors thank J. Swerts for performing the low-angle XRD
measurements.
\end{acknowledgments}


\pagebreak

\begin{thebibliography}{00}

\bibitem{Ohno}
H.~Ohno, A.~Shen, F.~Matsukura, A.~Oiwa, A.~Endo, S.~Katsumoto,
and Y.~Iye, Appl.~Phys.~Lett. {\bf69}, 363 (1996).

\bibitem{Sanvito}
S.~Sanvito, G.~Theurich, N.~Hill, J.~Supercond. {\bf15}, 85
(2002).

\bibitem{Maca}
F.~M\'{a}ca, J.~Ma\v{s}ek, Phys.~Rev.~B {\bf65}, 235209 (2002).

\bibitem{Erwin}
S.~C.~Erwin, A.~G.~Petukhov, Phys.~Rev.~Lett. {\bf89}, 227201
(2002).

\bibitem{Hayashi}
T.~Hayashi, Y.~Hashimoto, S.~Katsumoto, and Y.~Iye,
Appl.~Phys.~Lett. {\bf78}, 1691 (2001).

\bibitem{Potashnik1}
S.~J.~Potashnik, K.~C.~Ku, S.~H.~Chun, J.~J.~Berry, N.~Samarth,
and P.~Schiffer, Appl.~Phys.~Lett. {\bf79}, 1495 (2001).

\bibitem{Potashnik2}
S.~J.~Potashnik, K.~C.~Ku, R.~Mahendiran, S.~H.~Chun, R.F.~Wang,
N.~Samarth, and P.~Schiffer, Phys.~Rev.~B {\bf66}, 012408 (2002).

\bibitem{Kuryliszyn}
I.~Kuryliszyn, T.~Wojtowicz, X.~Liu, J.~K. Furdyna,
W.~Dobrowolski, J.-M.~Broto, M.~Goiran, O.~Portugall, H.~Rakoto,
and B.~Raquet, Acta~Phys.~Pol.,~A {\bf102}, 659 (2002).

\bibitem{Edmonds1}
K.~W.~Edmonds, K.~Y.~Wang, R.~P.~Campion, A.~C.~Neumann,
N.~R.~S.~Farley, B.~L.~Gallagher, and C.~T.~Foxon,
Appl.~Phys.~Lett. {\bf81}, 4991 (2002).

\bibitem{Ku}
K.~C.~Ku, S.J.~Potashnik, R.~F.~Wang, S.~H.~Chun, P.~Schiffer,
N.~Samarth, M.~J.~Seong, A.~Mascarenhas, E.~Johnston-Halperin,
R.~C.~Meyers, A.~C.~Gossard and D.~D.~Awschalom, Appl.~Phys.~Lett.
{\bf82}, 2302 (2003).

\bibitem{Chiba}
D.~Chiba, K.~Takamura, F.~Matsukura, and H.~Ohno,
Appl.~Phys.~Lett. {\bf82}, 3020 (2003).

\bibitem{Wang}
K.~Y.~Wang, R.~P.~Campion, K.~W.~Edmonds, M.~Sawicki, T.~Dietl,
C.~T.~Foxon and B.~L.~Gallagher, presented at the 27th
International Conference on Physics of Semiconductors, Flagstaff,
Arizona, USA, July 2004, (AIP Proceedings)

\bibitem{VanEsch}
A.~Van Esch, L.~Van Bockstal, J.~De Boeck, G.~Verbanck, A.~S.~Van
Steenbergen, P.~J.~Wellman, B.~Grietens, R.~Bogaerts, F.~Herlach,
and G.~Borghs, Phys. Rev. B {\bf56}, 13103 (1997).

\bibitem{Yu}
K.~M.~Yu, W.~Walukiewicz, T.~Wojtowicz, I.~Kuryliszyn, X.~Liu,
Y.~Sasaki, and J.~K. Furdyna, Phys.~Rev.~B {\bf65}, 201303 (2002).

\bibitem{Edmonds2}
K.~W.~Edmonds, P.~Boguslawski, K.~Y.~Wang, R.~P.~Campion,
S.~N.~Novikov, N.~R.~S.~Farley, B.~L.~Gallagher, C.~T.~Foxon,
M.~Sawicki, T.~Dietl, M.~Buongiorno~Nardelli, and J.~Bernholc,
Phys.~Rev.~Lett. {\bf92}, 037201 (2004).

\bibitem{Stone}
M.~B.~Stone, K.~C.~Ku, S.~J.~Potashnik, B.~L.~Sheu, N.~Samarth,
and P.~Schiffer, Appl.~Phys.~Lett. {\bf83}, 4568 (2003).

\bibitem{Brandt}
M.~S.~Brandt, S.~T.~B.~Goennenwein, T.~A.~Wassner, F.~Kohl,
A.~Lehner, H.~Huebl, T.~Graf, M.~Stutzmann, A.~Koeder, W.~Schoch,
and A.~Waag, Appl.~Phys.~Lett. {\bf84}, 2277 (2004).

\bibitem{Dietl}
T.~Dietl, H.~Ohno, and F.~Matsukura, Phys.~Rev.~B {\bf63}, 195205
(2001).

\bibitem{Sawicki}
M.~Sawicki, F.~Matsukura, T.~Dietl, G.~M.~Schott, C.~Ruester,
G.~Schmidt, L.~W.~Molenkamp, and G.~Karczewski, J.~Supercon.
{\bf16}, 7 (2003).

\bibitem{Fukumura}
T.~Fukumura, T.~Shono , K.~Inaba , T.~Hasegawa , H.~Koinuma,
F.~Matsukura and H.~Ohno, Physica E {\bf10}, 135 (2001).

\bibitem{Welp}
U.~Welp, V.~K.~Vlasko-Vlasov, X.~Liu, J.~K.~Furdyna and
T.~Wojtowicz, Phys.~Rev.~Lett. {\bf90}, 167206 (2003).

\bibitem{Tang}
H.~X.~Tang, R.~K.~Kawakami, D.~D.~Awschalom, and M.~L.~Roukes,
Phys.~Rev.~Lett. {\bf90}, 107201 (2003).

\bibitem{DeBoeck}
J.~De Boeck, R.~Oesterholt, A.~Van Esch, H.~Bender,
C.~Bruynseraede, C.~Van Hoof and G.~Borghs, Appl. Phys. Lett.
{\bf68}, 2744 (1996).


\end{thebibliography}
\end{document}